\documentclass[12pt]{article}
\usepackage[letterpaper, left=1in,top=1in,right=1in,bottom=1in]{geometry}
\usepackage[T1]{fontenc} 

\usepackage{amssymb}
\usepackage{pifont}

\usepackage{etoc} 
\usepackage{color}
\usepackage{tikz}
\usetikzlibrary{shapes,decorations,arrows,calc,arrows.meta,fit,positioning}
\tikzset{
    -Latex,auto,node distance =1 cm and 1 cm,semithick,
    state/.style ={ellipse, draw, minimum width = 0.7 cm},
    point/.style = {circle, draw, inner sep=0.04cm,fill,node contents={}},
    bidirected/.style={Latex-Latex,dashed},
    el/.style = {inner sep=2pt, align=left, sloped},
    roundnode/.style={circle, draw=green!60, fill=green!5, very thick, minimum size=7mm},
    squarednode/.style={rectangle, draw=blue!60, fill=blue!5, very thick, minimum size=5mm},
}
\definecolor{red}{HTML}{DB4437}
\definecolor{blue}{HTML}{4285F4}
\definecolor{green}{HTML}{0F9D58}
\definecolor{yellow}{HTML}{F4B400}

\usepackage{makecell}

\usepackage{CJKutf8} 
\usepackage{mathptmx} 

\usepackage{sectsty}
\sectionfont{\centering}

\usepackage[style=apa,backend=biber,isbn=false,uniquename=false,doi=true,url=false,eprint=false,dashed=false,annotation=false]{biblatex}
\AtEveryBibitem{%
  \clearname{translator}%
  \clearlist{language}%
  \clearfield{note}%
}
\bibliography{./MSDF.bib}

\usepackage{hyperref}
\usepackage{longtable}
\usepackage{multirow}
\usepackage{amsmath}
\mathchardef\mhyphen="2D 

\usepackage{todonotes}
\usepackage{color}
\usepackage{threeparttable}
\usepackage{tabularx}
\usepackage{rotating}
\usepackage{tablefootnote}

\usepackage{siunitx}
\usepackage{titlesec}
\titleformat*{\subsubsection}{\normalsize\itshape}
\usepackage[all]{nowidow}

\usepackage{setspace}
\makeatletter
\newcommand\iraggedright{%
  \let\\\@centercr\@rightskip\@flushglue \rightskip\@rightskip
  \leftskip\z@skip}
\makeatother
\iraggedright

\usepackage[capposition=top]{floatrow}

\usepackage{arydshln} 
\usepackage{appendix}
\usepackage{textcomp}
\usepackage{subcaption}
\usepackage{color,soul}
\usepackage{enumitem} 

\usepackage{ntheorem}
\theoremseparator{:}
\theoremheaderfont{\kern-1cm\normalfont\bfseries}

\usepackage[normalem]{ulem}
\useunder{\uline}{\ul}{}

\usepackage{graphicx}
\graphicspath{{./figs/}}

\usepackage{booktabs}

\newcolumntype{L}[1]{>{\raggedright\let\newline\\\arraybackslash\hspace{0pt}}m{#1}}
\newcolumntype{C}[1]{>{\centering\let\newline\\\arraybackslash\hspace{0pt}}m{#1}}
\newcolumntype{R}[1]{>{\raggedleft\let\newline\\\arraybackslash\hspace{0pt}}m{#1}}

\usepackage{tikz}
\usepackage{pgfplots}
\pgfplotsset{compat=newest}

\usepackage{listings}
\usepackage{xcolor}

\colorlet{punct}{red!60!black}
\definecolor{background}{HTML}{EEEEEE}
\definecolor{delim}{RGB}{20,105,176}
\colorlet{numb}{magenta!60!black}

\def\backtick{\char18}
\lstdefinelanguage{json}{
    basicstyle=\linespread{1}\footnotesize\raggedright\noindent,
    numbers=left,
    numberstyle=\scriptsize,
    stepnumber=1,
    numbersep=8pt,
    showstringspaces=false,
    breaklines=true,
    frame=lines,
    backgroundcolor=\color{background},
    literate=
     *{\{}{{{\color{delim}{\{}}}}{1}
      {\}}{{{\color{delim}{\}}}}}{1}
      {[}{{{\color{delim}{[}}}}{1}
      {]}{{{\color{delim}{]}}}}{1}
      {`}{{{\color{punct}{\backtick}}}}{1},
}

\begin{document}

\thispagestyle{empty}
\newcommand{\mytitle}{Integrating Transparent Models, LLMs, and Practitioner‑in‑the‑Loop:\\A Case of Nonprofit Program Evaluation}

\newcommand{\myabstract}{Public and nonprofit organizations often hesitate to adopt AI tools because most models are opaque even though standard approaches typically analyze aggregate patterns rather than offering actionable, case-level guidance. This study tests a practitioner-in-the-loop workflow that pairs transparent decision-tree models with large language models (LLMs) to improve predictive accuracy, interpretability, and the generation of practical insights. Using data from an ongoing college-success program, we build interpretable decision trees to surface key predictors. We then provide each tree's structure to an LLM, enabling it to reproduce case-level predictions grounded in the transparent models. Practitioners participate throughout feature engineering, model design, explanation review, and usability assessment, ensuring that field expertise informs the analysis at every stage. Results show that integrating transparent models, LLMs, and practitioner input yields accurate, trustworthy, and actionable case-level evaluations, offering a viable pathway for responsible AI adoption in the public and nonprofit sectors. \emph{(145 words)}}

\begin{titlepage}
    \thispagestyle{empty}

    \begin{center}
        \vspace*{-1cm}

        \singlespacing\textsc{\mytitle}\\~\\

        {Ji \uppercase{Ma}} \\
        \linespread{1}\small{LBJ School of Public Affairs, The University of Texas at Austin\\Gradel Institute of Charity, University of Oxford}\\~\\

        {Albert \uppercase{Casella}} \\
        \linespread{1}\small{Michael \& Susan Dell Foundation}\\~\\

    \end{center}

    \begingroup
    \onehalfspacing

    \begin{abstract}
        \noindent{\myabstract}
    \end{abstract}


    \endgroup

    \vspace{0.5cm}

    \noindent\rule{4cm}{0.4pt}\\
    \noindent\linespread{1}\footnotesize{\emph{Correspondence}: JM, \href{mailto:maji@austin.utexas.edu}{maji@austin.utexas.edu}; AC, \href{mailto:albert.casella@dell.org}{albert.casella@dell.org}. \emph{Funding}: The project is partly supported by (1) Michael \& Susan Dell Foundation, (2) Academic Development Funds from the RGK Center, and computing resources through (3) the Texas Advanced Computing Center at UT Austin \autocite{KeaheyLessonsLearnedChameleon2020} and (4) Dell Technologies (Client Memory Team and AI Initiative PoC Lead Engineer Wente Xiong). \emph{Compliance with Ethical Standards}: The author declares that this study complies with required ethical standards. \emph{Conflict of Interest}: The author declares no known conflict of interest.}

\end{titlepage}

\begingroup
\onehalfspacing
\normalsize
\etocdepthtag.toc{mtchapter}
\etocsettagdepth{mtchapter}{subsection}
\etocsettagdepth{mtappendix}{none}
\setcounter{tocdepth}{2}


\endgroup

\clearpage
\setcounter{page}{1}
\section{Introduction}

Predicting and monitoring project outcomes at the individual case level is critical for effective management, targeted interventions, and robust evaluation in the public and nonprofit sectors \autocite{FineProgramEvaluationPractice2000}. Case-level analyses enable organizations to identify specific cases needing attention, allocate resources efficiently, and intervene promptly to prevent negative outcomes. However, traditional social scientists' approaches to project evaluation have often focused on explanatory modeling at an aggregate level, seeking to understand underlying mechanisms rather than on predicting individual outcomes \autocite{HofmanIntegratingExplanationPrediction2021}. This focus is valuable for building theories and informing program strategy, but often overlooks case-specific nuances and actionable insights for practitioners. Scholars and practitioners both often struggle to translate findings from explanatory models into specific, practical strategies for case management at the individual level.

Recently, scholars have shown increased interest in integrating explanatory and predictive modeling approaches to address these limitations. Machine learning (ML) and artificial intelligence (AI) methods have begun to permeate the nonprofit sector, enhancing predictive accuracy and providing actionable insights. For instance, \textcite{YangMachineLearningApproachUnderstanding2025} employed random forest and gradient boosting algorithms to predict various performance dimensions (e.g., financial stability, service quality) of Canadian nonprofit sport organizations, revealing complex non-linear relationships and improving explanatory power beyond traditional linear regression. Similarly, \textcite{HesseUsingMachineLearning2025} applied random forest models to identify peer donors most likely to transition into organizational donors, achieving high predictive accuracy and informing targeted donor retention strategies.

Despite these advancements, two key challenges remain critical in order to apply academic research and advanced technologies in the nonprofit sector. (1) Models need to generate interpretable, actionable insights at the individual case level because practitioners require clear, specific guidance to effectively intervene. Without clear and actionable interpretations, predictive insights remain abstract and difficult to operationalize. (2) Predictive models need to ensure transparency because complex ML techniques often produce opaque decision-making processes despite achieving a high level of accuracy. This opacity could undermine the trust and acceptance of practitioners, particularly those who do not have many experiences with interpreting ML models. This may then limit their willingness to rely on model-driven decisions in sensitive and high-stake contexts, as has been discussed by \textcite{RudinStopExplainingBlack2019}.

In response to these challenges, this study integrates transparent predictive modeling \autocite{RudinStopExplainingBlack2019} and case-level interpretation within a practitioner-in-the-loop workflow \autocite{SalipanteManagersKnowledgeGenerators2003}, to identify students in a college scholarship program focused on who are at risk of not graduating on time (Section \ref{sec:data_source}). We first develop transparent decision-tree models to identify critical predictors and establish baseline predictive performance (Section \ref{sec:decision_tree}). We then use large language models (LLMs) to generate clear, case-specific explanations in natural language, which are directly informed by the transparent decision-tree models (Section \ref{sec:llm_interpretation}). Throughout the process, practitioners actively engaged in feature selection, model improvement, prompt engineering, and interpretation validation to ensure the analysis was consistently aligned with practical expertise and organizational needs (Section \ref{sec:practitioner_in_the_loop}; Figure \ref{fig:practitioner_in_the_loop_workflow}). Our results demonstrate that this integrated approach achieves strong predictive accuracy, and that augmenting LLM-generated explanations with a curated program knowledge base significantly improves their perceived safety and fairness, offering a responsible and practical model for AI adoption in the public and nonprofit sectors.

\section{Methods}

\subsection{Data Source} \label{sec:data_source}

The data originates from a longstanding scholarship program established in 2004 to support promising students with financial need and a determination to pursue higher education. The program systematically collects data through structured surveys administered to students each academic semester. These surveys gather data on student demographics, academic performance, financial circumstances, and personal challenges the student may be facing. Academic performance and student financial aid data are verified using university documentation. Responses are linearly aggregated into a simple risk score for each student, serving as a key predictor of whether the student is likely to need program assistance to graduate within four years of first enrollment (i.e. ``on time''). Students identified as ``at-risk'' trigger interventions by project staff, who provide tailored support and resources to address the student's specific needs. Currently, this system is managed by a small team case managers responsible for approximately 2,000 students. Given current staff capacity and a need to optimize program efficiency, the project requires enhanced predictive capabilities to identify at-risk students and enable more customized intervention plans.

While the dataset collected via surveys is comprehensive and accurate, it presents significant challenges for predictive modeling due to its highly imbalanced nature---approximately 75\% of students in the dataset used for this study graduate within the targeted timeframe. Consequently, the primary monitoring goal of this project is to effectively identify students at risk of delayed graduation and provide personalized, case-level intervention strategies.

\subsection{Measures}

\textit{Outcome variable}. The primary outcome is a binary indicator of on-time graduation, defined as completing a bachelor’s degree within four years of first enrollment. This measure aligns with the program’s core mission of supporting timely completion for its students.

\textit{Predictors}. Drawing on prior studies \autocite{PageMoreDollarsScholars2019,PageImprovingCollegeAccess2016,HopeBoostFirstgenerationLowincome2016,KehoeBridgingCollegeCompletion2017,SahadewoEssaysEconomicsEducation2017} and meetings with project case managers (Section \ref{sec:practitioner_in_the_loop}), we selected predictors associated with graduation outcomes in theory and in practice. These predictors fall into five thematic blocks: (1) demographic background, (2) pre-college academic profile, (3) time-varying academic progress, (4) financial circumstances, and (5) institutional context. \autoref{tab:predictors} lists every predictor, its survey label, and the expected directional relationship with on-time graduation.

\begin{table}[htbp]
    \centering
    \small
    \caption{\textsc{Predictors of On-Time Graduation}}
    \label{tab:predictors}
    \begin{tabularx}{\textwidth}{l l l l}
        \hline\hline
        ~~~ & \textbf{Variable (survey label)}                              & \textbf{Data type}$^{a}$ & \textbf{Expected effect$^{b}$} \\
        \hline
        \multicolumn{4}{l}{\textit{A. Demographic \& Background (static)}}                                                              \\
            & gender                                                        & Categorical              & mixed                          \\
            & scholar\_ethnicity, scholar\_race                             & Categorical              & mixed                          \\
            & citizenship                                                   & Categorical              & +                              \\
            & has\_children                                                 & Categorical              & --                             \\
            & numberotherdependents                                         & Numeric                  & --                             \\
        \addlinespace
        \multicolumn{4}{l}{\textit{B. Pre-college Academic Profile (static)}}                                                           \\
            & highschoolgpa\_pct                                            & Numeric                  & +                              \\
            & readeracademicscore, readertotalscore                         & Numeric                  & +                              \\
            & finalacademicscore, finaltotalscore                           & Numeric                  & +                              \\
        \addlinespace
        \multicolumn{4}{l}{\textit{C. Academic Progress (panel)}}                                                                       \\
            & gpacumulativecurrent                                          & Numeric                  & +                              \\
            & hoursattempted, hourscompleted                                & Numeric                  & +                              \\
            & creditstowardsdegree                                          & Numeric                  & +                              \\
            & totalcreditsneeded                                            & Numeric                  & --                             \\
            & enrollment (Full Time, Not Enrolled, Part Time)               & Categorical              & +                              \\
            & changed\_enrollment\_type                                     & Categorical              & --                             \\
        \addlinespace
        \multicolumn{4}{l}{\textit{D. Financial Circumstances (panel)}}                                                                 \\
            & costofattendance                                              & Numeric                  & --                             \\
            & efcamount (Expected Family Contribution)                      & Numeric                  & +                              \\
            & grantaid                                                      & Numeric                  & +                              \\
            & loanamountoffered, loanamountaccepted                         & Numeric                  & mixed                          \\
            & totalloandebt                                                 & Numeric                  & --                             \\
        \addlinespace
        \multicolumn{4}{l}{\textit{E. Institutional Context (panel)}}                                                                   \\
            & collegesector (Public, Private Nonprofit, Private For-Profit) & Categorical              & mixed                          \\
            & collegetype (2-yr, 4-yr, Mixed)                               & Categorical              & mixed                          \\
            & areaofstudy (various)                                         & Categorical              & mixed                          \\
            & persistrateyoy\_any (Persist YOY, any enrollment)             & Categorical              & +                              \\
            & persistrateyoy\_ft2ft (Persist YOY, full-time only)           & Categorical              & +                              \\

        \hline\hline
    \end{tabularx}
    \begin{flushleft}
        \footnotesize $^{a}$Numerical values are converted to percentiles to reflect the position among peers.\\$^{b}$``+'' = higher value likely increases probability of graduating on time; ``--'' = lowers probability; ``mixed'' = theoretically ambiguous or context-dependent.
    \end{flushleft}
\end{table}

\subsection{Predictive and Explanatory Models}

A distinct goal of this work was to better understand how the predictive model and interpretation can be effectively integrated into the workflow of a case manager. Thus, model selection did not necessarily focus on benchmarking various ML models, but rather testing the validity and utility of an integrated approach. To enhance predictive accuracy while preserving transparency, we implemented a decision-tree model because it is designed to clearly illustrate how input variables contribute to a final classification. For practical case-level decision-making, we integrate an LLM (i.e., \texttt{GPT-o3}) to convert model outputs into accessible, natural-language explanations tailored explicitly for case managers. The refinement and finalization of these models were embedded into a practitioner-in-the-loop workflow, ensuring that domain experts actively contributed throughout feature engineering, model design, and results interpretation.

\subsubsection{Predictive Model with Transparency: Decision Tree} \label{sec:decision_tree}

Decision tree models provide visual representations of how input features guide classification decisions. These models are inherently transparent, in that they explicitly identify key branching points based on feature thresholds to show the combinations of characteristics that place students at greater risk of delayed graduation. Compared ML models that may remain accurate but opaque (i.e. ``black-box models''), decision trees are equipped to foster practitioner understanding and trust, making them particularly suitable for applied settings such as nonprofit program monitoring. As \textcite{RudinStopExplainingBlack2019} argues, when decisions carry significant real-world implications, interpretable models may be more appropriate compared to post-hoc explanations of opaque algorithms (e.g., neural networks). Moreover, evidence suggests that interpretable models often achieve comparable or even superior predictive performance relative to black-box methods \autocite{ChenInterpretableModelGlobally2018}.

To select an optimal decision-tree model, we conducted a rigorous hyperparameter tuning process using a grid search approach (i.e., iterating all possible parametric combinations). Appendix \ref{sec:grid_search} includes details of this process.

\subsubsection{Case-Level Prediction and Interpretation: Large Language Model} \label{sec:llm_interpretation}

We instruct the LLM to generate clear, case-specific explanations based on the decision tree model with case managers as the intended audience. Prompt engineering is central in this process and includes: (1) the structure of the trained decision tree, (2) the specific input data of the student, (3) the model's prediction for the student, (4) the key predictors driving the prediction, and (5) potential intervention strategies to address areas of concern. This approach was structured to enable case managers to quickly interpret the results, identify critical issues, and effectively respond with student-specific interventions.

To establish a performance baseline, we also tested whether an LLM could predict a student's risk status directly from their data, without relying on the decision tree model or program-specific knowledge. This \emph{LLM-zero-shot} approach \autocite{KojimaLargeLanguageModels2022,WeiFinetunedLanguageModels2022} serves as a benchmark to evaluate the added value of our integrated decision-tree method. As shown in Table \ref{tab:model_performance} and Figure \ref{fig:auc_roc_llm}, the decision tree model consistently outperforms the LLM-zero-shot approach across all cohort years on all performance metrics. This performance gap underscores the critical role of decision tree in improving predictive accuracy for this task.

\subsection{Evaluation}

\subsubsection{Model Performance Metrics} \label{sec:model_performance_metrics}

To assess the predictive quality of the decision tree models, we adopted a suite of metrics that are robust to class imbalance. In this case, a naive model could achieve high overall accuracy by always predicting the majority class due to imbalance. In such cases, additional metrics are needed, specifically \textit{precision}, \textit{recall}, \textit{F1-score}, and area under the receiver operating characteristic curve (\textit{AUC-ROC}). Each of these metrics are defined below:

\begin{itemize}[noitemsep, topsep=0pt, partopsep=0pt, parsep=0pt]
    \item \textit{Precision} quantifies the proportion of correctly identified at-risk students among all students flagged by the model, reflecting how actionable the alerts are for case managers.
    \item \textit{Recall} measures the share of truly at-risk students that the model successfully identifies, capturing the model's ability to minimize missed interventions.
    \item \textit{F1-score} is the harmonic mean of precision and recall, providing a single summary that balances the costs of false positives and false negatives in an imbalanced setting.
    \item \textit{AUC-ROC (Area under the Receiver Operating Characteristic Curve)} evaluates discriminatory power across all possible classification thresholds, offering a threshold-independent view of model performance.
\end{itemize}

\subsubsection{Practitioner-in-the-Loop} \label{sec:practitioner_in_the_loop}

\begin{figure}[htbp]
    \centering
    \caption{\textsc{Practitioner-in-the-Loop Workflow}}
    \label{fig:practitioner_in_the_loop_workflow}
    \includegraphics[width=0.9\textwidth]{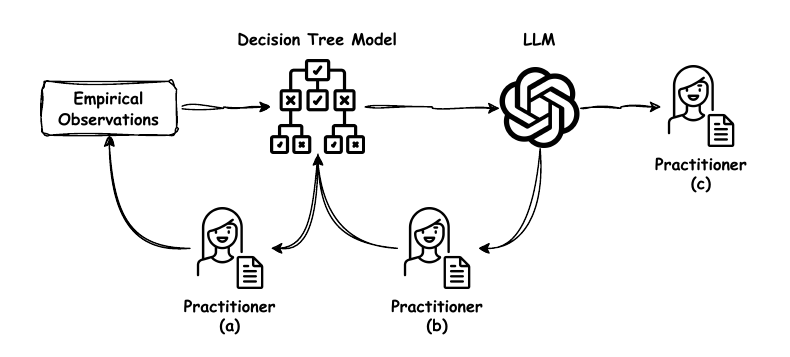}
\end{figure}

Quantitative scores alone are insufficient for deployment in a high-stakes, student-facing context, as the decision tree model output may not be practically useful for case managers and other front-line practitioners. In this project, we embedded a ``practitioner-in-the-loop'' approach (Figure \ref{fig:practitioner_in_the_loop_workflow}) at every major stage (research design, model validation, prompt engineering, and results interpretation) to help ensure that model results are appropriate. This approach treated practitioners as both knowledge generators and ultimate consumers, an approach advocated for in the nonprofit and research translation literature \autocites[144, Table 2]{SalipanteManagersKnowledgeGenerators2003}{BonneyCanCitizenScience2016}.

\textit{Feature and Framing Review} (Figure \ref{fig:practitioner_in_the_loop_workflow}\emph{a}): Adopting an abductive perspective \autocite{TaylorLeapingRealWorldRelevance2018}, the predictors used in the decision tree models were identified in collaboration with front-line staff. Program managers participated in multiple working sessions to review candidate predictors for face validity and potential bias. This step ensured that model inputs consistently aligned with both theoretical frameworks and practical experience.

\textit{Prompt Engineering Cycles} (Figure \ref{fig:practitioner_in_the_loop_workflow}\emph{b}): After the decision tree was finalized, we conducted iterative prompt-engineering sessions with case managers. Each session provided the LLM with: (1) the tree's rule path for a given student, (2) the student's data points, and (3) the model's risk prediction. Case managers critiqued the generated explanations on clarity, relevance, and usability, leading to successive refinement of the prompt template.

\textit{Usability Assessment} (Figure \ref{fig:practitioner_in_the_loop_workflow}\emph{c}): In a final session, three case managers rated 30 anonymized LLM explanations on a 5-point Likert scale covering \textit{usefulness}, \textit{transparency}, and \textit{safety} \autocites{WangUnderstandingUserExperience2024}{CalvanoLeveragingLargeLanguage2025}{QuttainahCostUsabilityCredibility2024}{HaoExploringCollaborativeDecisionmaking2024}. Appendix \ref{sec:usability_assessment} and \ref{sec:usability_questionnaire} provide more details.

Together, the quantitative performance metrics and qualitative feedback from practitioners provide a comprehensive evaluation of the modeling pipeline.

\section{Results}

\subsection{Descriptive Statistics}

The panel is unbalanced and comprises 2,245 students followed across multiple years. The sample is comprised of students from multiple cohorts. Most (75.06\%) students in the sample graduated within four years of first enrollment. \autoref{tab:descriptive_statistics} summarizes the descriptive statistics of key panel variables that consistently appear in the final decision trees of all cohort years. Cost of attendance averages hover around \$34 k with substantial variability (SD $\approx$ \$18--22 k) and a maximum of \$128 k. Credits toward degree mirror a student's academic progress, climbing sharply from a mean of 33.6 credits after the student's first year to a mean of 114.5 credits in the student's fourth year. Cumulative GPA remains remarkably stable after the student's first year, centering on 3.26/4 for the first three years and dipping only slightly to 3.25/4 in the 4th year. Grant aid grows steadily from \$11.8k to \$17.2k, suggesting increased financial support as students advance, whereas total loan debt shows low means ($\approx$ \$1.4 k) and medians of zero each year, indicating that most students either incur no debt or rely primarily on grants and the scholarship provided by the program. Together, these figures depict a trajectory characterized by rising academic progress and grant aid with stable academic performance and limited reliance on loans.

\begin{table}[htbp]
    \centering
    \small
    \caption{\textsc{Key Descriptive Statistics by Cohort Year (Years 1--4)}}
    \label{tab:descriptive_statistics}
    \begin{tabular}{lrrrrrr}
        \hline\hline
        Cohort & Count  & Mean       & SD         & Min  & 50\%       & Max         \\
        \hline
        \multicolumn{7}{l}{\textit{Cost of attendance}}                             \\
        1      & 2\,233 & 34\,391.94 & 18\,408.63 & 0.00 & 27\,860.00 & 90\,828.00  \\
        2      & 2\,206 & 33\,345.29 & 19\,678.21 & 0.00 & 27\,472.50 & 87\,688.00  \\
        3      & 2\,106 & 33\,886.89 & 20\,493.87 & 0.00 & 27\,915.50 & 109\,405.00 \\
        4      & 1\,867 & 34\,045.06 & 22\,172.49 & 0.00 & 28\,394.00 & 128\,334.00 \\
        \addlinespace
        \multicolumn{7}{l}{\textit{Credits toward degree}}                          \\
        1      & 2\,220 & 33.58      & 22.93      & 0.00 & 29.00      & 180.00      \\
        2      & 2\,189 & 55.13      & 28.03      & 0.00 & 51.00      & 205.00      \\
        3      & 2\,075 & 85.08      & 33.65      & 0.00 & 83.00      & 255.00      \\
        4      & 1\,852 & 114.48     & 39.33      & 0.00 & 120.00     & 420.00      \\
        \addlinespace
        \multicolumn{7}{l}{\textit{GPA (cumulative)}}                               \\
        1      & 2\,245 & 3.260      & 0.536      & 0.44 & 3.370      & 4.00        \\
        2      & 2\,236 & 3.264      & 0.530      & 0.50 & 3.380      & 4.00        \\
        3      & 2\,201 & 3.267      & 0.522      & 0.69 & 3.370      & 4.00        \\
        4      & 2\,011 & 3.252      & 0.527      & 0.69 & 3.360      & 4.00        \\
        \addlinespace
        \multicolumn{7}{l}{\textit{Grant aid}}                                      \\
        1      & 2\,233 & 11\,781.55 & 7\,296.92  & 0.00 & 10\,920.00 & 95\,291.00  \\
        2      & 2\,206 & 14\,245.07 & 12\,273.95 & 0.00 & 11\,495.00 & 89\,621.00  \\
        3      & 2\,106 & 15\,546.40 & 14\,958.32 & 0.00 & 12\,086.00 & 108\,761.00 \\
        4      & 1\,867 & 17\,172.65 & 18\,355.00 & 0.00 & 12\,009.00 & 93\,495.00  \\
        \addlinespace
        \multicolumn{7}{l}{\textit{Total loan debt}}                                \\
        1      & 1\,449 & 1\,382.15  & 4\,608.48  & 0.00 & 0.00       & 60\,684.00  \\
        2      & 1\,449 & 1\,382.15  & 4\,608.48  & 0.00 & 0.00       & 60\,684.00  \\
        3      & 1\,424 & 1\,351.59  & 4\,572.04  & 0.00 & 0.00       & 60\,684.00  \\
        4      & 1\,265 & 1\,380.36  & 4\,691.60  & 0.00 & 0.00       & 60\,684.00  \\
        \hline\hline
    \end{tabular}
\end{table}

\subsection{Model Performance}

As reported in \autoref{tab:model_performance}, the decision trees deliver strong and consistent classification results across cohort years. Overall accuracy ranges between 0.88 and 0.90 across cohort years. Precision for the minority ``at-risk'' class is high (0.78--0.86), while recall spans 0.68--0.73, yielding F1-scores between 0.74 and 0.78. In contrast, the majority ``graduate on time'' class demonstrates very high precision (0.90--0.92) and recall (0.94--0.96), producing F1-scores at or above 0.92. These patterns reflect the roughly 3:1 class imbalance: the models minimize false positives without severely sacrificing sensitivity.

\begin{sidewaysfigure}[htbp]
    \centering
    \begin{subfigure}[b]{0.49\textwidth}
        \centering
        \includegraphics[width=\textwidth]{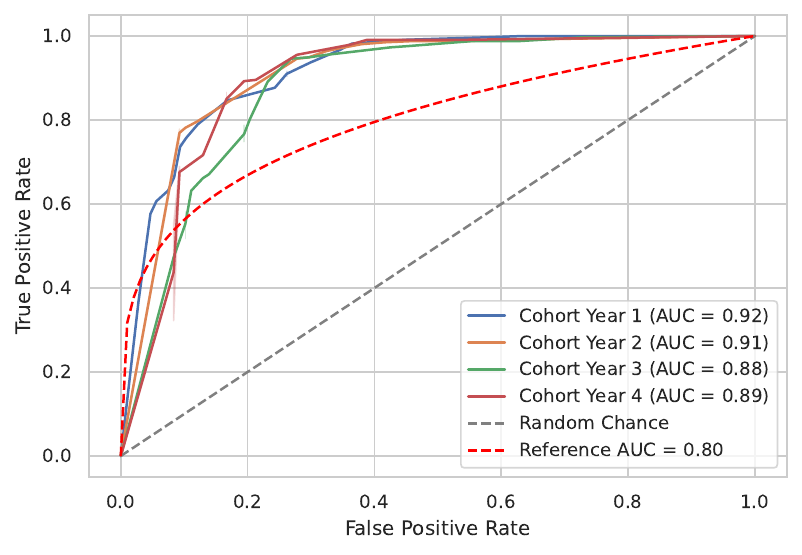}
        \caption{Decision Tree}
        \label{fig:auc_roc_dt}
    \end{subfigure}
    \hfill
    \begin{subfigure}[b]{0.49\textwidth}
        \centering
        \includegraphics[width=\textwidth]{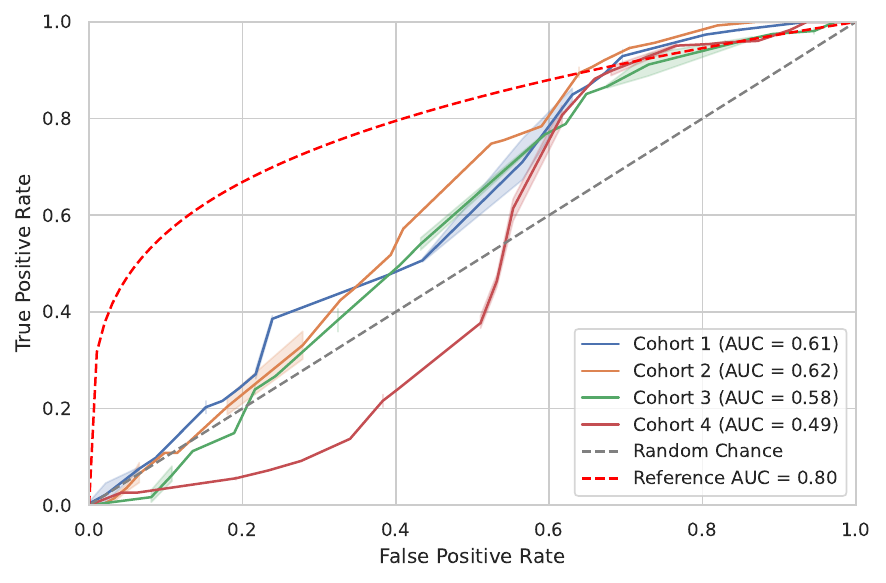}
        \caption{LLM-zero-shot (\texttt{GPT-o3})}
        \label{fig:auc_roc_llm}
    \end{subfigure}
    \caption{\textsc{AUC-ROC Curves for Decision Tree and LLM-zero-shot Models Across Cohorts}}
    \label{fig:auc_roc}
\end{sidewaysfigure}

Discriminatory power remains robust across cohort years, as illustrated by the ROC curves in \autoref{fig:auc_roc}. The corresponding AUC values are 0.92 (cohort year 1), 0.91 (cohort year 2), 0.88 (cohort year 3), and 0.89 (cohort year 4), all comfortably surpassing the 0.80 benchmark for acceptable classification \autocite{CorbaciogluReceiverOperatingCharacteristic2023} and well above random chance (i.e., 0.5). The modest dip for cohort year 3 aligns with its slightly lower AUC, yet the model still attains acceptable F1-score (0.77) and AUC (0.80).

These metrics indicate that the transparent decision-tree model provides reliable early-warning signals while keeping false alarms at a manageable level. This reliability is essential for practitioner workflows with limited intervention bandwidth. The visualizations of each tree structure are available at \href{https://osf.io/kfpej/?view_only=709ea028b2b0433c9c8ed2ed034c5c58}{Open Science Framework},\footnote{\url{https://osf.io/kfpej/?view_only=709ea028b2b0433c9c8ed2ed034c5c58}} where users can explore the exact splits that underpin each prediction.

\begin{table}[htbp]
    \centering
    \small
    \caption{\textsc{Model Performance Metrics by Cohort Year}}
    \label{tab:model_performance}
    \begin{tabular}{crcccc}
        \hline\hline
        \textbf{Cohort Year} & \textbf{Prediction}        & \textbf{Precision} & \textbf{Recall} & \textbf{F1-score} & \textbf{\#Case} \\
        \hline
        \multirow{2}{*}{1}
                             & At-risk                    & 0.78 (0.41)        & 0.70 (0.72)     & 0.74 (0.52)       & 107 (46)        \\
                             & Graduate on time           & 0.91 (0.89)        & 0.94 (0.69)     & 0.92 (0.78)       & 333 (153)       \\[0.5ex]
                             & \textit{Weighted Accuracy} &                    &                 & 0.88 (0.72)       & 440 (199)       \\
        \hline
        \multirow{2}{*}{2}
                             & At-risk                    & 0.86 (0.55)        & 0.68 (0.84)     & 0.76 (0.66)       & 108 (61)        \\
                             & Graduate on time           & 0.90 (0.91)        & 0.96 (0.70)     & 0.93 (0.79)       & 334 (139)       \\[0.5ex]
                             & \textit{Weighted Accuracy} &                    &                 & 0.89 (0.75)       & 442 (200)       \\
        \hline
        \multirow{2}{*}{3}
                             & At-risk                    & 0.81 (0.31)        & 0.73 (0.87)     & 0.77 (0.46)       & 108 (37)        \\
                             & Graduate on time           & 0.92 (0.95)        & 0.95 (0.57)     & 0.93 (0.71)       & 334 (161)       \\[0.5ex]
                             & \textit{Weighted Accuracy} &                    &                 & 0.89 (0.66)       & 442 (198)       \\
        \hline
        \multirow{2}{*}{4}
                             & At-risk                    & 0.84 (0.35)        & 0.72 (0.85)     & 0.78 (0.50)       & 108 (47)        \\
                             & Graduate on time           & 0.91 (0.92)        & 0.96 (0.52)     & 0.93 (0.67)       & 335 (153)       \\[0.5ex]
                             & \textit{Weighted Accuracy} &                    &                 & 0.90 (0.63)       & 443 (200)       \\
        \hline\hline
    \end{tabular}
    \begin{flushleft}
        \footnotesize \emph{Note}: Values in parentheses show the performance of the baseline \emph{LLM-zero-shot} (\texttt{GPT-o3}) model, which predicts outcomes directly from student data without using the decision tree or program-specific knowledge.
    \end{flushleft}
\end{table}

\subsection{Case‐Level Prediction and Interpretation} \label{sec:case_level}

To translate model scores into actionable guidance, we employed an LLM which ingests the trained decision‐tree, the focal student's data, and the predicted classification. The model was instructed to then produce a plain-language explanation targeted at case managers. The prompt template (Appendix \ref{sec:llm_prompts}) instructed the LLM to (1) state the predicted class with its node-level probability, (2) walk through every split in the path using conversational comparisons (e.g., ``GPA below 10th percentile''), (3) highlight the top five drivers in everyday terms, (4) flag borderline features that could flip the outcome, and (5) close with concise, adviser-oriented takeaways.

To leverage organizational expertise and enhance explanation quality, we experimented with two prompt variants: one that provides only the decision-tree path and student data (Appendix Figure \ref{fig:llm_prompt_wo_kb}), and another that supplements this information with curated program knowledge distilled from case managers' best practices (Appendix Figure \ref{fig:llm_prompt_wt_kb}). The program knowledge encompasses established intervention strategies, including academic tutoring protocols, financial counseling frameworks, and mental health support resources, which the LLM can reference to generate more contextually appropriate and actionable recommendations. This in-context learning approach \autocite{DongSurveyIncontextLearning2024} has been shown to enable LLMs to produce explanations that align model predictions with organizational expertise.

Appendix Figure \ref{fig:llm_output} shows one anonymized output for a student in their third year with a predicted \textit{NoGrad4yr} classification at 83\% probability. The LLM first articulates the risk score, then lays out a four-node path that moves from ``zero loan debt''---a potential proxy for working long hours to avoid debt---to ``very low GPA,'' concluding in a terminal node that historically contains 25 non-graduates versus 5 graduates. The explanation pinpoints low GPA, high cost of attendance, and modest remaining credits as the dominant risk factors, but also notes that the GPA is just below an actionable threshold for academic tutoring. Importantly, the final bullet list converts these insights into concrete adviser actions (e.g., ``raise GPA one letter grade,'' ``explore additional grant aid or financial counseling to reduce cost pressure''), demonstrating how model logic is bridged to practice.

\subsection{Usability Assessment}

\begin{figure}[htbp]
    \centering
    \includegraphics[width=0.9\textwidth]{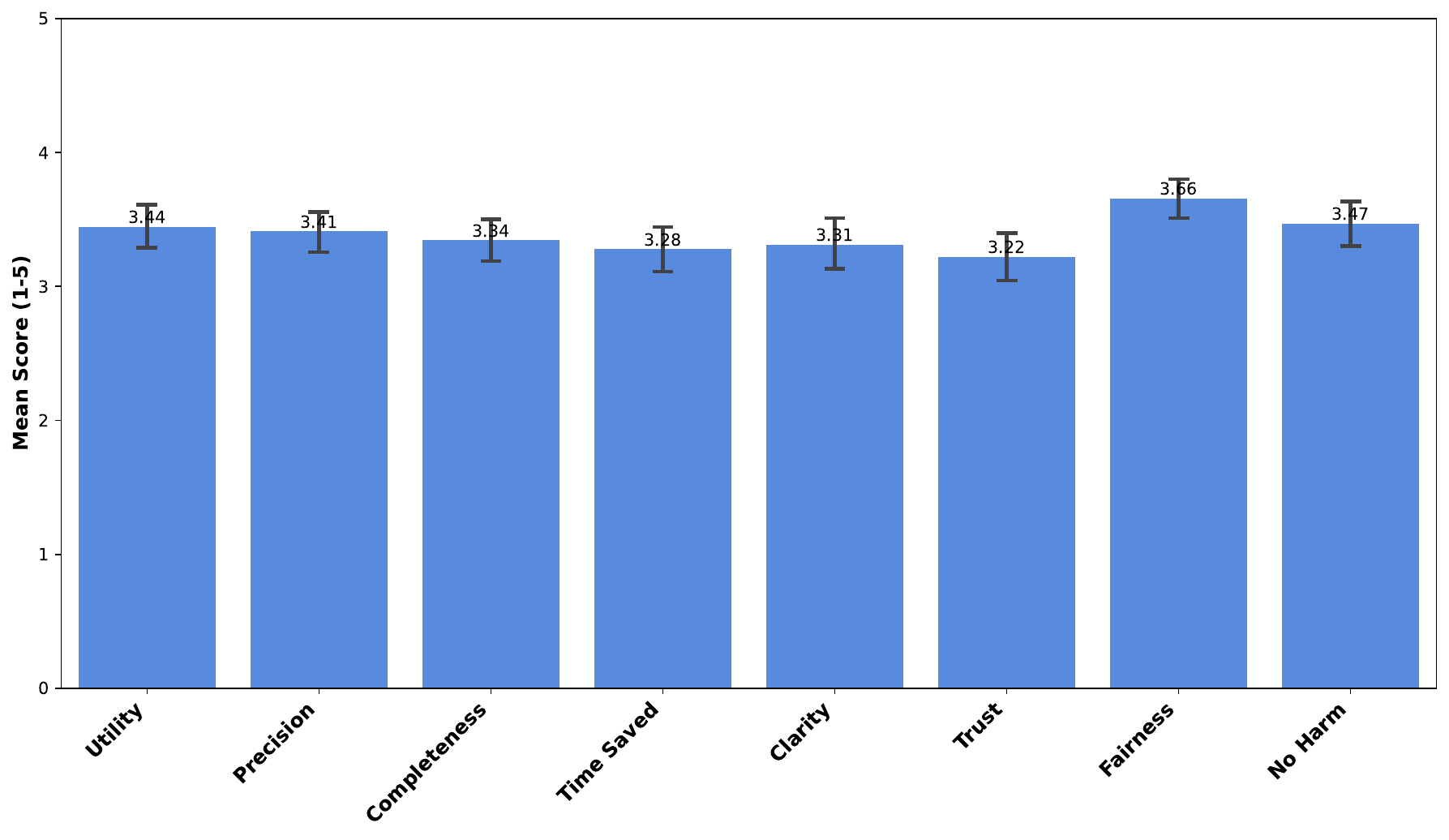}
    \caption{\textsc{Usability Test Results for LLM-Generated Explanations}}
    \label{fig:usability_test_results}
\end{figure}

Thirty LLM-generated at-risk predictions (e.g., Appendix Figure \ref{fig:llm_output}) were randomly selected from the full evaluation set, and rated by three case managers on a five-point Likert scale covering \emph{usefulness}, \emph{transparency}, and \emph{safety} (Appendix Table \ref{tab:usability_metrics}). Figure \ref{fig:usability_test_results} summarizes the results. The means of all dimensions exceed 3.0, suggesting that the explanations are generally perceived as useful, transparent, and safe. Notably, the scores of \emph{Fairness} ($Mean = 3.66, SD = 0.72$) and \emph{No Harm} ($Mean = 3.46, SD = 0.81$) received relatively higher ratings compared to other dimensions, suggesting that case managers perceive these explanations as less likely to introduce bias or cause negative consequences, both of which are critical considerations for high-stakes student interventions.

To understand whether incorporating program knowledge (Appendix Figure \ref{fig:llm_prompt_wt_kb}) improves explanation quality compared to using only the decision-tree path and student data (Appendix Figure \ref{fig:llm_prompt_wo_kb}), we conducted a controlled comparison. LLM-generated explanations using the two variants of prompts were provided to case managers without the case managers knowing whether a specific explanation produced from a prompt with program knowledge. Since each case manager evaluated multiple explanations and specific student cases might inherently be easier or harder to interpret, we employed regression analysis to isolate the effect of program knowledge while controlling for potential confounding factors. Specifically, we ran separate regressions using the Likert scores on each dimension as outcome variables and a binary indicator of prompt type (with vs. without program knowledge) as the key predictor, controlling for fixed effects: case manager ID, student case ID, and cohort year, with standard errors clustered by case manager ID.

\begin{figure}[htbp]
    \centering
    \includegraphics[width=.65\textwidth]{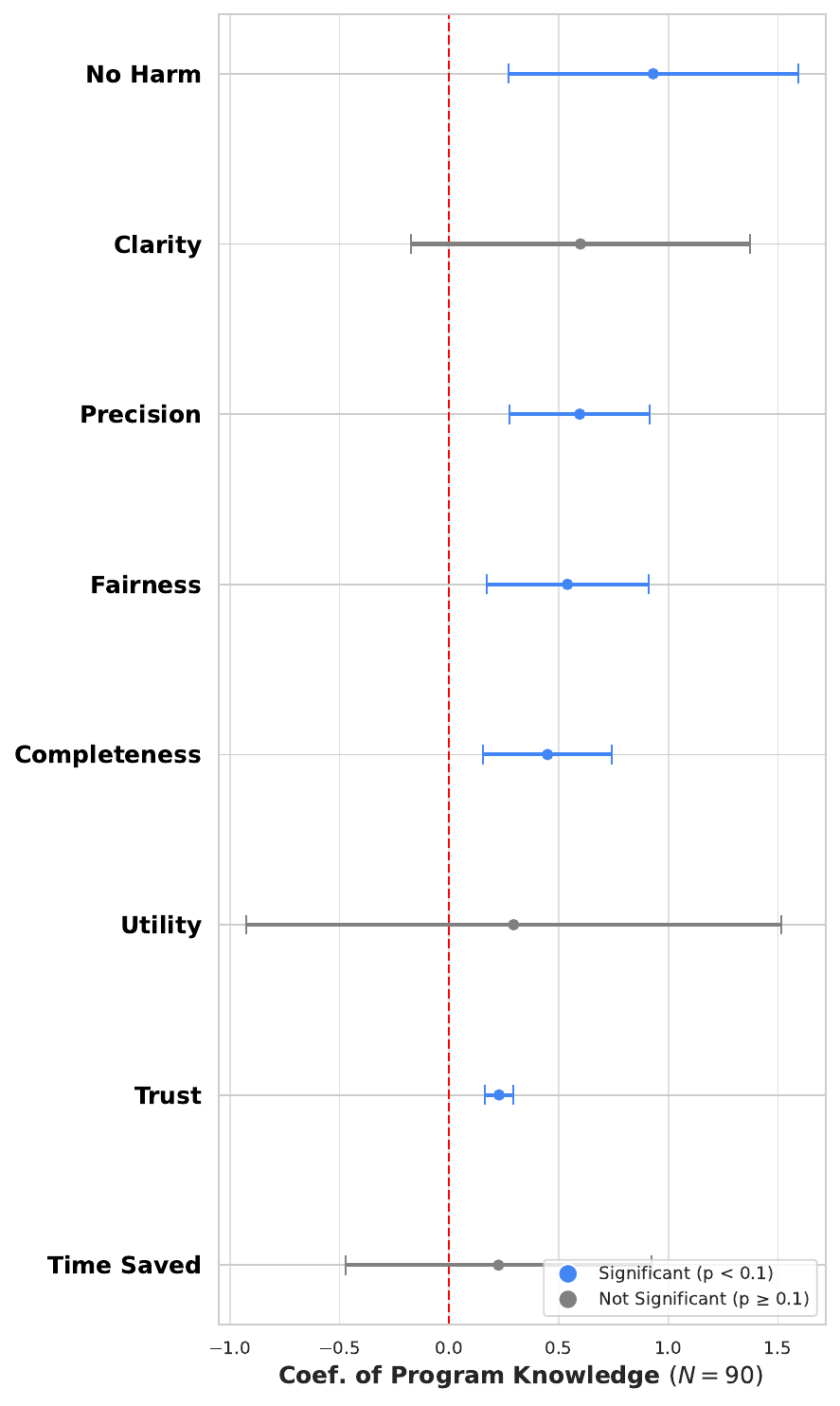}
    \caption{\textsc{Usability Test Results: With vs. Without Program Knowledge}}
    \label{fig:dsp_survey_regression_results}
    \begin{flushleft}
        \footnotesize \emph{Note}: Each point represents the estimated coefficient for ``with program knowledge'' (base group: without program knowledge) from separate regressions on each usability dimension (i.e., Y-axis = DVs). Error bars indicate 90\% confidence intervals.
    \end{flushleft}
\end{figure}

Figure \ref{fig:dsp_survey_regression_results} shows the regression results. Incorporating program knowledge significantly improved most dimensions, with the exception of ``Clarity,'' ``Utility,'' and ``Time Saved.'' The largest positive effect is on ``No Harm'' ($\beta = 0.93, p = 0.02$), indicating that explanations with program knowledge are rated nearly a full point higher on the five-point scale for safety. This is followed by ``Precision'' ($\beta = 0.60, p < 0.01$) and ``Fairness'' ($\beta = 0.54, p = 0.02$), both showing substantial improvements of over half a scale point. These findings suggest that domain expertise enhances the trustworthiness and safety of AI-generated explanations rather than their operational efficiency.

Interestingly, feeding program knowledge does not appear to improve perceived time savings, perceived utility, or clarity. This pattern aligns with our practitioner-in-the-loop philosophy: while domain knowledge may not always accelerate decision-making or simplify explanations, it critically enhances the safety and fairness dimensions that practitioners value most when deploying AI in sensitive educational contexts. The results underscore that responsible AI adoption in the nonprofit sector requires not just technical accuracy but also the integration of contextual expertise to ensure ethical and trustworthy deployment.

\section{Discussion}

This study offers an example of integrating transparent predictive modeling with AI interpretation for case-level decision-making in a nonprofit scholarship program. We developed a decision-tree model to predict which students are at risk of delayed graduation and paired it with an LLM to produce actionable explanations for case managers. This approach addresses two common challenges in nonprofit analytics: the need for interpretable, actionable insights at the individual level and the need for trustworthy, transparent models in high-stakes contexts. We found that a simple, interpretable model provided sufficient predictive accuracy. When combined with an LLM, the model produced clear insights at the case-level. There are several lessons from this work that nonprofit organizations adopting AI may wish to consider.

\paragraph{Workflow with AI tools integrated.} A first lesson regards the importance of defining the role of AI within a project's workflow, including the division of labor between AI and human practitioners. In this case, we integrated AI as a decision-support tool with a practitioner-in-the-loop framework integrated at onset. A decision tree model was intentionally chosen to maximize transparency, and the model remained highly accurate in its predictions. In our implementation, the AI components \textit{inform} decisions, rather than direct them: the model flags at-risk students, and the LLM provides reasoning and suggestions, but case managers remain the ultimate decision-makers. This delineation ensures the appropriate use of AI while practitioners managed nuanced judgment and intervention planning. Further, case managers understood at the onset that the AI system was there to assist their expertise. This approach may facilitate buy-in and trust among staff new to using AI, as well as those wary of using it in a systematic way for decision-making. This case supports recent calls from \textcite{ChenInterpretableModelGlobally2018} and \textcite{YangMachineLearningApproachUnderstanding2025} to consider practitioners as collaborators in the analytic process.

\paragraph{Knowledge base for AI, grounded in human expertise.} A second lesson centers on the value of establishing a \emph{knowledge base} grounded in existing program expertise. Integrating staff knowledge into the model improved the relevance and actionability of LLM recommendations. When the LLM had access to program-specific knowledge, its explanations were rated significantly higher by case managers on dimensions of \textit{fairness}, \textit{safety}, and \textit{trust} compared to explanations without that knowledge base. In practice, this meant the LLM suggestions aligned with empiric intervention strategies and avoided potentially insensitive or infeasible recommendations. Organizations interested in integrating LLM into their workflow may benefit from intentionally incorporating organizational knowledge into the model, as this helps ensure that insights align with context-specific reality and reinforces practitioner trust in model outputs.

\paragraph{Transparency over complexity.} Results support the use of transparent models when accuracy is sufficient and staff utility is a priority. These models can be sufficiently accurate and may be more appropriate than more sophisticated ensemble ``black-box'' models when the purpose pertains to high-stakes predictive tasks \autocites{ChenInterpretableModelGlobally2018}{RudinStopExplainingBlack2019}. Despite its simple structure, the model achieved strong predictive performance while offering clear logic for each prediction. For example, instead of simply telling a case manager that ``Student Y has a 83\% chance of not graduating on time,'' the model might explain:

\begin{quotation}

    \noindent\textit{``Student Y is at high risk because their GPA dropped below 2.5 this term while taking on extra work hours; similar cases who reduced work hours experienced improved academic outcomes. Consider adjusting this student’s workload via increased financial support or providing tutoring.''}

\end{quotation}

In our case, the decision tree's performance was within a few points of more complex algorithms used in similar contexts \autocite{YangMachineLearningApproachUnderstanding2025}, offering little practical incentive to incur the opacity of a more sophisticated approach. If a simpler model can achieve the needed accuracy, organizations should consider opting for transparency due to the potential dividends in staff acceptance and clarity. Moreover, using an interpretable model simplifies the task of diagnosing errors or biases, which is crucial for responsible AI use in social sector projects.


\clearpage
\begingroup
\singlespacing
\sloppy
\printbibliography[title={References}]
\endgroup

\clearpage
\begingroup
\section*{\textsc{Online Appendix}}
\begin{appendix}
    \begin{refsection}
        \renewcommand\thetable{\Alph{section}\arabic{table}}
\renewcommand\thefigure{\Alph{section}\arabic{figure}}
\setcounter{footnote}{0}
\setcounter{page}{1}

\begin{center}
    \vspace*{-1cm}

    \singlespacing\textsc{\mytitle}\\~\\

    {Ji \uppercase{Ma}} \\
    \linespread{1}\small{The University of Texas at Austin}\\~\\

    {Albert \uppercase{Casella}} \\
    \linespread{1}\small{Michael \& Susan Dell Foundation}\\~\\

\end{center}

\vspace{-2cm}

\begingroup
\onehalfspacing
\normalsize
\etocdepthtag.toc{mtappendix}
\etocsettagdepth{mtchapter}{none}
\etocsettagdepth{mtappendix}{subsubsection}
\setcounter{tocdepth}{3}
\tableofcontents
\endgroup

\clearpage

\section{Methods}

\subsection{Grid Search for Hyperparameter Tuning} \label{sec:grid_search}

We explored three critical hyperparameters known to influence decision-tree complexity and performance:

\begin{itemize}[noitemsep, topsep=0pt, partopsep=0pt, parsep=0pt]
    \item \textit{Criterion}: The method used to measure the quality of a split, tested across ``gini,'' ``entropy,'' and ``log\_loss.''.
    \item \textit{Maximum depth}: The maximum allowable depth of the tree, ranging from 1 to 29.
    \item \textit{Minimum samples per leaf}: The minimum number of samples required to be at a leaf node, ranging from 5 to 29.
\end{itemize}

Using \textit{four-fold cross-validation}, we systematically evaluated combinations of these hyperparameters. Cross-validation splits the dataset into four segments, repeatedly training on three segments and validating on the remaining one. This technique helps to robustly evaluate model performance on unseen data and prevents overfitting, ensuring the model generalizes well beyond the training dataset.

To identify the best-performing hyperparameters, we optimized specifically for the weighted F1-score (Section \ref{sec:model_performance_metrics}). This choice of metric accounts effectively for class imbalance, emphasizing both precision and recall for each class while giving appropriate consideration to their proportions.

\subsection{Usability Assessment} \label{sec:usability_assessment}

Hallucination and long-context processing are two of the most important concerns for users of LLMs \autocite[12]{WangUnderstandingUserExperience2024}. In empirical studies of human-AI collaboration, researchers consistently focus on three key dimensions: \textit{usefulness}, \textit{transparency}, and \textit{safety} \autocites[79]{CalvanoLeveragingLargeLanguage2025}[7]{QuttainahCostUsabilityCredibility2024}{HaoExploringCollaborativeDecisionmaking2024}. Based on these studies, we defined specific metrics for usability evaluation as Table \ref{tab:usability_metrics} shows.

\begin{table}[htbp]
    \centering
    \caption{\textsc{Usability Evaluation Dimensions and Metrics}}
    \label{tab:usability_metrics}
    \begin{tabularx}{\textwidth}{l l X}
        \hline\hline
        \textbf{Dimension}             & \textbf{Metric} & \textbf{Definition}                                                                                                                              \\
        \hline
        \multirow{8}{*}{\emph{Usefulness}}
                                       & Utility         & Relevance and practicality of the results for case-manager tasks.                                                                                \\
                                       & Precision       & Clarity and appropriate detail of explanations provided by the model.                                                                            \\
                                       & Completeness    & Extent to which the explanations fully address key aspects necessary for decision-making and intervention planning.                              \\
                                       & Time Saved      & Amount of time saved by using model-generated explanations compared to manual interpretation and analysis.                                       \\
        \hline
        \multirow{4}{*}{\emph{Transparency}}
                                       & Clarity         & Ease with which case managers can comprehend how the model reached its conclusions.                                                              \\
                                       & Trust           & Case managers' confidence in the model-generated explanations and overall accuracy of predictions.                                               \\
        \hline
        \multirow{5}{*}{\emph{Safety}} & Fairness        & Absence of unfair discrimination in model recommendations and explanations.                                                                      \\
                                       & No Harm         & Absence of harmful recommendations that could lead to negative consequences for students, such as inappropriate interventions or stigmatization. \\
        \hline\hline
    \end{tabularx}
\end{table}

\clearpage
\section{LLM Prompts for Case-Level Explanations} \label{sec:llm_prompts}

\begin{figure}[htbp]
    \centering
    \includegraphics[width=1\textwidth]{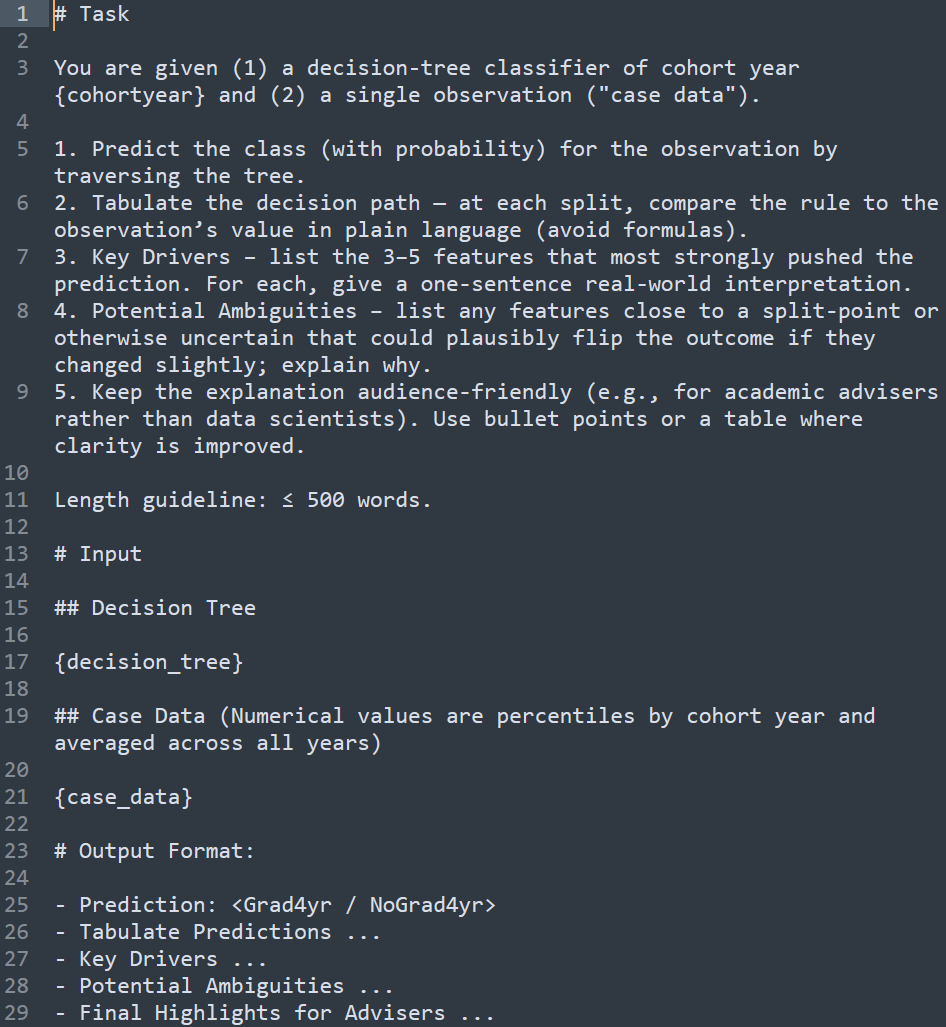}
    \caption{\textsc{LLM Prompt Template for Case-Level Explanations (Without Knowledge Base)}}
    \label{fig:llm_prompt_wo_kb}
\end{figure}

\begin{figure}[htbp]
    \centering
    \includegraphics[width=1\textwidth]{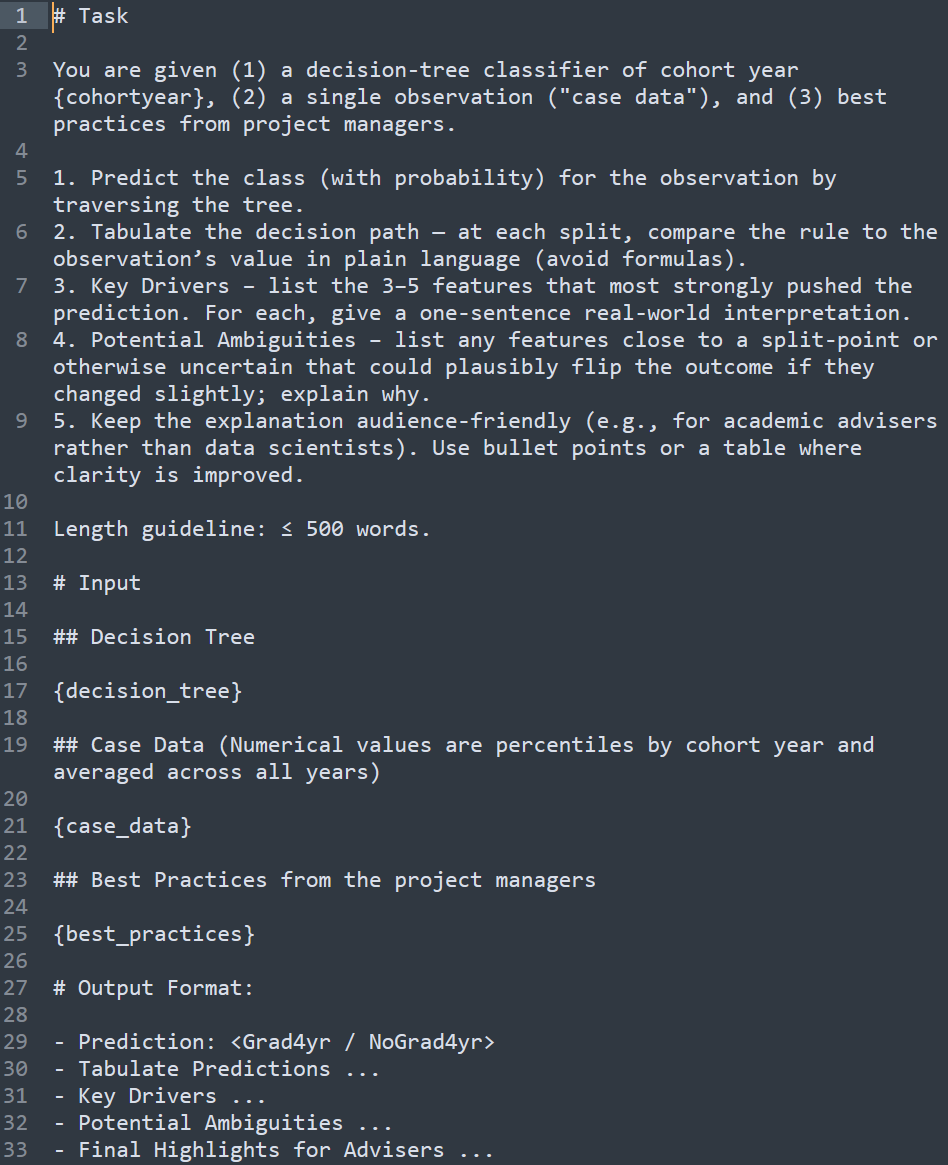}
    \caption{\textsc{LLM Prompt Template for Case-Level Explanations (With Knowledge Base for In-Context Learning)}}
    \label{fig:llm_prompt_wt_kb}
\end{figure}

\clearpage
\begingroup
\singlespacing
\setlength{\parindent}{0pt}

\section{Usability Assessment Questionnaire} \label{sec:usability_questionnaire}
\thispagestyle{empty}

Thank you for participating in our usability evaluation. This usability test is part of a broader project integrating transparent predictive models, large language models, and practitioner expertise to help case managers identify students at risk of delayed graduation. The test specifically evaluates the \emph{usefulness}, \emph{transparency}, and \emph{safety} of AI-generated, case-level explanations derived from a decision-tree model. By systematically collecting feedback from case managers, we ensure that model outputs are not only accurate but also clear, trustworthy, and actionable for real-world, student-facing decision-making. For each explanation provided by the AI model, please rate its quality on the criteria below. Your honest feedback is critical for improving our predictive and interpretive models.

\begin{center}\
    \small
    \begin{tabular}{cl}
        \toprule
        \textbf{Score} & \textbf{Interpretation} \\
        \midrule
        1              & Strongly disagree       \\
        2              & Disagree                \\
        3              & Neutral                 \\
        4              & Agree                   \\
        5              & Strongly agree          \\
        \bottomrule
    \end{tabular}
\end{center}

\begin{table}[htbp]
    \centering
    \small
    \caption*{Questionnaire items used to evaluate each anonymized explanation.}
    \label{tab:questionnaire_items}
    \begin{tabular}{r p{0.50\linewidth}}
        \toprule
        \multicolumn{2}{l}{\textit{Usefulness}}                                                                                  \\ \addlinespace[2pt]
        Utility      & The explanation is relevant and practical for my tasks as a case manager.                                 \\[2pt]
        Precision    & The explanation is clear and provides an appropriate level of detail.                                     \\[2pt]
        Completeness & The explanation fully covers the information needed for decision-making and intervention planning.        \\[2pt]
        Time Saved   & Using this explanation significantly reduces the time required compared to manual interpretation.         \\[4pt]

        \multicolumn{2}{l}{\textit{Transparency}}                                                                                \\ \addlinespace[2pt]
        Clarity      & I can easily comprehend how the model reached its conclusions.                                            \\[2pt]
        Trust        & I am confident in the accuracy and reliability of the model-generated explanation.                        \\[4pt]

        \multicolumn{2}{l}{\textit{Safety}}                                                                                      \\ \addlinespace[2pt]
        Fairness     & The explanation is free from unfair discrimination or bias toward the student.                            \\[2pt]
        No Harm      & The explanation does not suggest any actions or interpretations that could negatively impact the student. \\
        \bottomrule
    \end{tabular}
\end{table}

\textbf{Additional Comments (optional).} Please provide any further suggestions or observations to enhance the model's performance or usability.

\endgroup

\clearpage
\thispagestyle{empty}
\begin{figure}[htbp]
    \centering
    \includegraphics[width=1\textwidth]{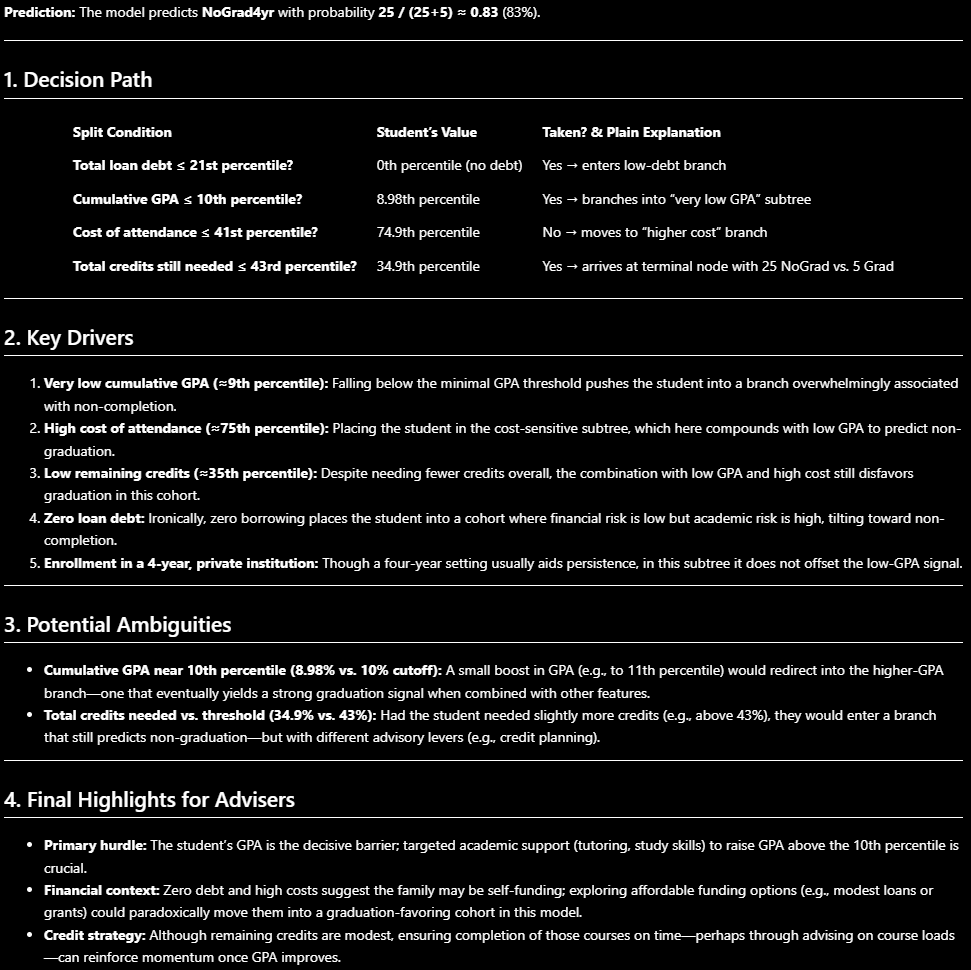}
    \caption{\textsc{Sample LLM Output for Case-Level Explanations}}
    \label{fig:llm_output}
\end{figure}

\clearpage
\thispagestyle{empty}
\begin{sidewaysfigure}[htbp]
    \centering
    \includegraphics[width=1\textwidth]{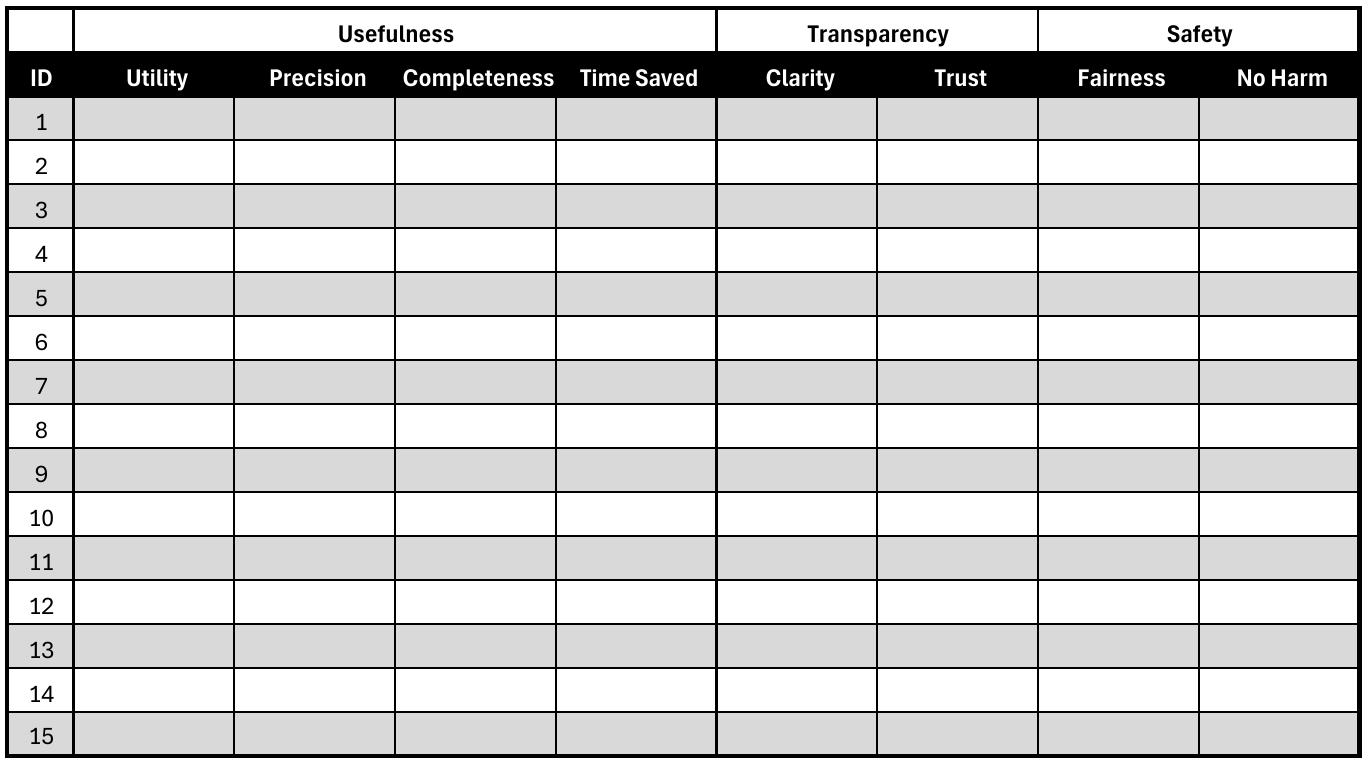}
    \caption*{Scoring Sheet for Usability Assessment / \textbf{Respondent ID: (\underline{\hspace{2cm}})}}
    \label{fig:scoring_sheet_1}
    \textbf{1: Strongly DISAGREE / 2: DISAGREE / 3: Neutral / 4: AGREE / 5: Strongly AGREE}
\end{sidewaysfigure}

\clearpage
\thispagestyle{empty}
\begin{sidewaysfigure}[htbp]
    \centering
    \includegraphics[width=1\textwidth]{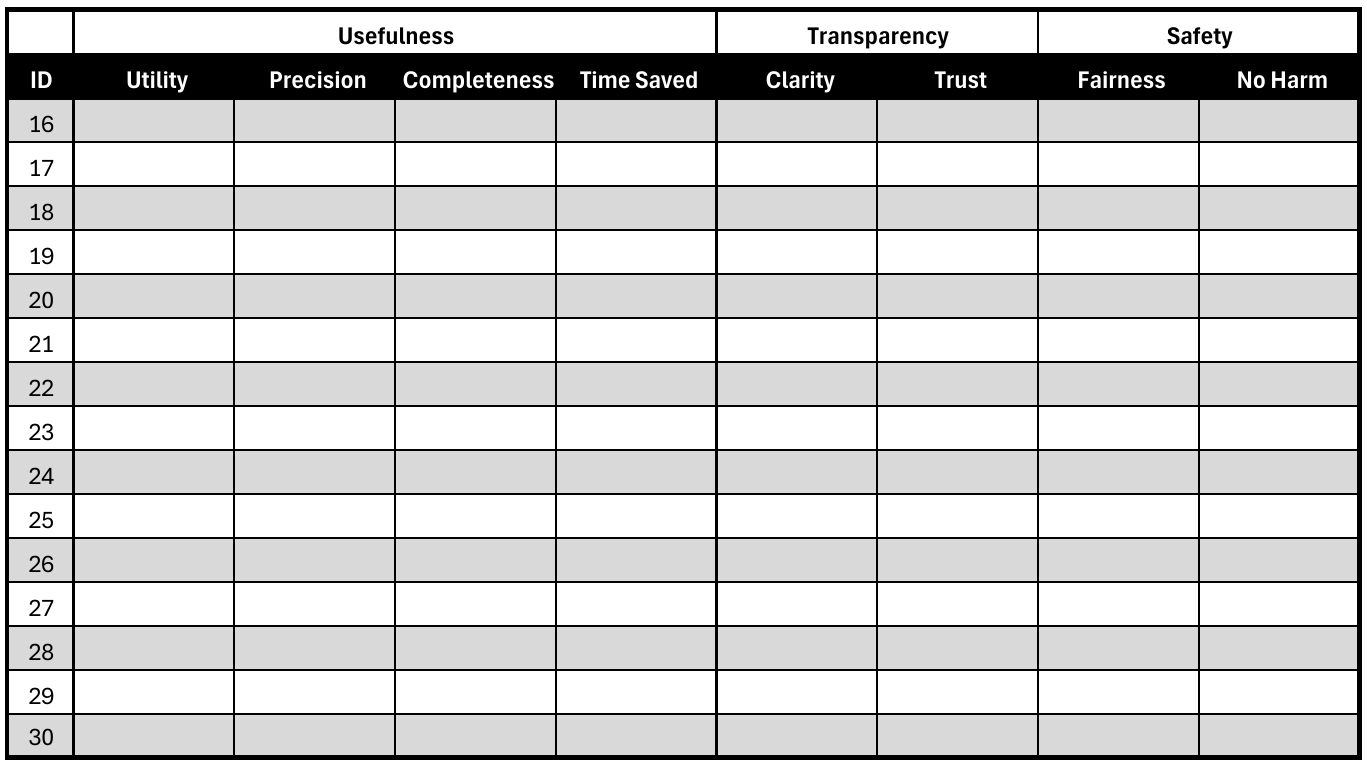}
    \caption*{Scoring Sheet for Usability Assessment / \textbf{Respondent ID: (\underline{\hspace{2cm}})}}
    \label{fig:scoring_sheet_2}
    \textbf{1: Strongly DISAGREE / 2: DISAGREE / 3: Neutral / 4: AGREE / 5: Strongly AGREE}
\end{sidewaysfigure}

\clearpage
\begingroup
\singlespacing
\sloppy
\printbibliography[title={References}]
\endgroup

    \end{refsection}
\end{appendix}
\endgroup

\end{document}